\documentclass[conference]{IEEEtran}
\IEEEoverridecommandlockouts
\usepackage{cite}
\usepackage{amsmath,amssymb,amsfonts}
\usepackage{algorithmic}
\usepackage{textcomp}
\usepackage{xcolor}

\usepackage{palatino}
\usepackage{graphicx}
\usepackage{float}
\usepackage{subfigure}
\usepackage{float}
\usepackage{color}
\usepackage{lipsum}
\usepackage{multicol}
\usepackage{multirow}
\usepackage{booktabs}

\def\BibTeX{{\rm B\kern-.05em{\sc i\kern-.025em b}\kern-.08em
    T\kern-.1667em\lower.7ex\hbox{E}\kern-.125emX}}
\begin{document}

\title{An Analysis of Pedestrians' Behavior in Emergency Evacuation Using Cellular Automata Simulation
\thanks{}
}

\author{\IEEEauthorblockN{Zhiguo Zhou, Huiyu Cai}
\IEEEauthorblockA{\textit{School of Remote Sensing and Information Engineering} \\
\textit{Wuhan University}\\
Wuhan, China \\
\{zhouzhikwo, chyqwegh\}@163.com}
\and
\IEEEauthorblockN{Zhixuan Zhou}
\IEEEauthorblockA{\textit{Hongyi Honor College} \\
\textit{Wuhan University}\\
Wuhan, China \\
kyriezoe@whu.edu.cn}
}

\maketitle

\begin{abstract}
To minimize property loss and death count in terror attacks and other emergent scenarios, attention given to timely and effective evacuation cannot be enough. 
Due to limited evacuation resources, i.e., number of available exits, there exists interdependence among pedestrians such as cooperation, competition and herd effect. Thus human factors---more specifically, pedestrians' behavior in emergency evacuation---play a significant role in evacuation research. Effective evacuation can only be reached when route planning are considered in conjunction with psychological dynamics, which is often ignored. In this paper, we analyze the effect of pedestrians' behavior, i.e., herd effect and knowledge of changing environment with Cellular Automata (CA) simulation. Results of the simulation show harmful effect of herd effect as well as highlight the importance of timely informing pedestrians of environmental change. Accordingly, we propose policy and procedural recommendations for emergency management of large, crowded structures.
\end{abstract}


\section{Introduction}

Terrorist forces are becoming more and more rampant, causing panic across the world. Fires are another major source of danger when people are gathered in a crowded structure. Thus establishing a robust and efficient plan for evacuating pedestrians is necessary.

Emergency evacuation plans include effective prevention and efficient handling confronting emergencies. Safeguarding personal and property security of pedestrians can make a difference to maintaining social stability. The design of emergency evacuation plans, as a vital part of the security system, has obtained more and more attention from relevant authorities and researchers. 

Generally, the goal of evacuation is to have all pedestrians leave the building as quickly and safely as possible. Upon notification of a required evacuation, individuals egress to and through an optimal exit in order to empty the building in the least amount of time. It has been investigated with various models, including macroscopic models and microscopic models. In macroscopic models, the interaction among pedestrians isn't considered in detail, so they are not suitable for evacuation in complex space. Microscopic models, on the contrary, overcome the above mentioned shortcoming and have drawn more attention from the research community. However, it is observed that many of them cannot be effectively applied to all kinds of emergencies and situations. From our perspective, this is partly due to lack of consideration of pedestrians' behavior during emergencies. 

Because of the limited evacuation resource, there often exists competition and cooperation between pedestrians, which resembles the classical prisoners' dilemma~\cite{Prisoner}, i.e., the optimal solution of population is inconsistent with that of individuals. In this paper we model competitive behavior in the simulation. Impact of herd effect and pedestrians' knowledge of the surroundings are analyzed using Cellular Automata simulation. 

The rest of our paper is organized as below. Section~\ref{sec:related work} surveys related work on human factors and evacuation models. Section~\ref{sec:preliminaries} gives our general assumptions and preliminary knowledge of cellular automata. In section~\ref{sec:model} baseline model and improved model for pedestrian dynamics analysis are proposed. Simulation results as well as corresponding implications are provided in Section~\ref{sec:evaluation}. Strategy recommendations and future work are discussed in Section~\ref{sec:discussion}.   

\section{Related Work}\label{sec:related work}

Evacuation planning has always been a hot research area and many models have been proposed. 

\cite{Metro Evacuation} gives a model for evaluation in subway fire, which emphasizes the number and width of escape exits. The authors believe that a subway station should have at least two exits in different directions, and the wider the exits, the better. \cite{CA} presents a model based on the parallel computational tool of cellular automata capable of simulating the process of disembarking in a small airplane seat layout, in search of ways to make it faster and safer under normal evacuation conditions, as well as emergency scenarios. \cite{survey} provides a comprehensive survey of evacuation planning models.

Human Factors, i.e., how people behave in emergency evacuation scenarios, plays a significant role in decision making and evacuation actions.

In~\cite{Shiwakoti17} the likely behaviours of train passengers in an emergency evacuation are explored in the form of questionnaires. The results showed that respondents are more likely to be reactive (e.g., wait for instruction from station staff) rather than proactive (e.g., move to exit); more likely to be cooperative (e.g., helping other people) than competitive (e.g., push other passengers); more likely to show herding or symmetry behaviour (e.g., following other passengers) than symmetric breaking behaviour (e.g., choose least crowded exit); and less likely to use the lifts, escalators and tunnels to escape in an emergency situation. In terms of demographic differences in behaviours, it's demonstrated that there are significant differences in the evacuation behaviours between males and females. Males are less likely to use emergency call buttons, call the emergency phone number or wait at the assembly area. Also, they are more likely to display competitive behaviour, choose the least crowded exit, and use the lifts, escalators and tunnels to escape in the event of an emergency evacuation. In contrast,  differences in behaviours are not as obvious among different age groups.

\cite{Manley16} emphasizes the need for greater participation from individuals with disabilities during the evacuation
planning process. Understanding and addressing the psychological needs of individuals with disabilities during emergency evacuations will help ensure the safety of everyone.

\cite{CA2} proposes a behavior-based cellular automaton (CA) model for pedestrian dynamics and show that for a given range of emergency degree, the enlarged emergency degree can shorten the evacuation time yet depresses cooperation enthusiasm. On the contrary, as dependence (familiarity) of pedestrians increases, the evacuation time and the fraction of cooperation increase.

In~\cite{Louvreanalysis} the authors analyze pedestrians' sequential movements, the spatial layout, and the relationship between them in the Louvre Museum using anonymous data collected through noninvasive Bluetooth sensors, which enable them to unveil some features of pedestrian behavior and spatial impact that shed some light on the mechanisms of museum overcrowding. 

\section{Preliminaries}\label{sec:preliminaries}

\subsection{General Assumptions}
\begin{figure*}[h]
\centering
\subfigure[Before evacuation]{
\begin{minipage}[t]{0.5\linewidth}
\centering
\includegraphics[scale=0.8]{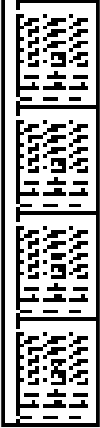}
\end{minipage}%
}%
\subfigure[After evacuation]{
\begin{minipage}[t]{0.5\linewidth}
\centering
\includegraphics[scale=0.8]{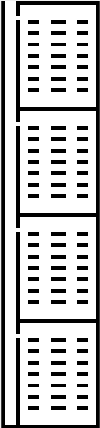}
\end{minipage}%
}%
\centering
\caption{Initial and terminate states.}
\label{comparison}
\end{figure*}

Without generality, we keep to the assumptions below in the whole model.
\begin{itemize}
    
    \item The exhibition halls and corridors are both rectangular; there are four exhibition halls on each floor, and each exhibition hall has the same layout; adjacent exhibition halls share a wall; the length of the corridor is the sum of internal lengths of the four exhibition halls and the thickness of the walls; exits of the exhibition halls are set on the same side, and exit of the corridor is set on one end of it. The layout is portrayed in Figure~\ref{comparison}.
    \item Group tourists and free tourists are both counted as individuals.
    \item During the evacuation process, nobody falls in a faint, has a heart attack or shows other symptoms; nor do we consider the effect of people's emotion changes on the acceptance of instructions.
    \item Fire, explosion, shooting and other situations requiring evacuation are in our consideration.
\end{itemize}

\subsection{Cellular Automata}

A cellular automata~\cite{cellular} consists of a regular grid of cells, each in one of a finite number of states, such as on and off. Each cell has its own risk, depending on its distance from the exit and the presence of obstacles around it. The grid can be in any finite number of dimensions. For each cell, a set of cells called its neighborhood is defined relative to the specified cell. An initial state (time $t = 0$) is selected by assigning a state for each cell. A new generation is created (advancing $t$ by 1), according to some fixed rule that determines the new state of each cell in terms of the current state of the cell and the states of the cells in its neighborhood. Typically, the rule for updating the state of cells is the same for each cell and does not change over time, and is applied to the whole grid simultaneously~\cite{CAConclusion}.

\section{Pedestrian Dynamics Model}\label{sec:model}

The majority of previous works assume that 1) pedestrians have same level of knowledge of the surrounding environment, i.e., spatial accessibility, which is stationary; 2) there is no competition among pedestrians and no herd effect. But it's not usually the case in real scenarios. In fact, individual knowledge and observation of the surroundings differ greatly. During emergent events, some exits which are not often utilized in normal times are opened. Competition and herd effect may have harmful effect. These factors are worth paying attention to. Similar implication is demonstrated in~\cite{change}. 

Here we take several characteristics of pedestrian behavior mentioned above into consideration.

\subsection{Baseline Model}

Pedestrian dynamics in emergencies can be simulated with cellular automata. Since exhibition halls and corridors are rectangular, we view them as two-dimensional meshes. Thus we assume layout and situation of each floor as in Figure~\ref{comparison}.

In the exhibition hall, distribution of tourists is determined by that of exhibits. The white part denotes walking area, the black lumps denote exhibits and the black dots denote pedestrians. If we want to keep track of change of each cell, we need to set up rules for each tourist. 

As a common sense, people will surely move a shorter distance to the exit. When a cell is occupied by walls, exhibits or tourists, we consider distance to the exit as the minimal integer larger than the diagonal length of the hall. Values of each cell represent the distance from the exit. A matrix is thus obtained, determining risk of each cell. The smaller the element value is, the safer the cell.

Considering his Moore neighborhood~\cite{CA3}, each pedestrian only needs to consider the risk of his eight neighbor cells and his own at each time step. If there is a cell of unique minimal risk among eight others, pedestrians will obviously choose this cell as their next target. He will not move if the cell he's standing on is already local optimal at the moment. Because risk of the nine cells may have multiple minimal values, pedestrians will choose one of them with same probability. Additionally, there may be multiple pedestrians competing for one cell at the same time. In this case, we also assign the cell to them with same probability.

From the above description, we have concisely set up the rules of movement for each cell. Each time step allows only one pedestrian to move when there's a tie, thus allowing only one person to walk out of the exhibition hall to the corridor. 

In short, each person can only occupy one cell---the size of which is $0.4*0.4m^2$---and move in white areas. Each value assigned to cells corresponds to a risk degree, based on which pedestrians can determine their movement for the next time step. With observation of his Moore neighborhood~\cite{meth}, individuals can move in eight directions or remain stationary and at the same time update their observation. One move takes 0.25 second~\cite{statistic}.

Specific rules in the exhibition hall are as follows:
\begin{enumerate}
    \item Calculate the distance between the cell and the exit $S(x, y)$. We assume the position of the door is $(x_0,y_0)$.

\begin{equation}
S(x,y)=\left\{
\begin{array}{rcl}
\sqrt{(x-x_0)^2+(y-y_0)^2} && (available)\\
M && (occupied)\\
\end{array} \right.
\end{equation}
*In the formula, (x, y) is the coordinate of cells, $(x_0,y_0)$ is the coordinate of doors. M denotes diagonal length of the exhibition hall.  

    \item Calculate risk of each cell $A(x,y)$.
    
\begin{equation}
 A(x,y)=\left\{
\begin{array}{rcl}
0&&(grid~at~the~door)\\
S(x,y) && (otherwise)\\
\end{array} \right.
\end{equation}

    \item Compare risk value of each neighbor cell and select the smallest one; choose one of them as next target with same probability if there is a tie; stay still if already staying in a local optimal position. 
    \item If multiple individuals compete for an idle position at the same time, the system will select them with same probability.
    \item When pedestrians arrive at the exit of the exhibition hall, we define the risk value as that of the cell beside it. 
\end{enumerate}

When considering corridors, it is assumed that only three rows of people can simultaneously walk. Pedestrians will follow the same rules of walking as in the exhibition halls. That is to say, people will also move a shorter distance to the exit. The exit is set at one end as shown in the figures. 

For implementation, we build a matrix A of size $116*26$ to denote all four exhibition halls and the corridor. The first column represents the wall of the corridor. The second, third and fourth columns represent walking areas of the corridor. The fifth column represents the walls of the exhibition halls with one door each. Other columns represent the exhibition halls. When all matrix elements representing the corridor become 0, all tourists have successfully evacuate. Figure~\ref{comparison} shows the states before evacuation and after respectively.

\subsection{Improved Model}
For better effectiveness and efficiency of evacuation, pedestrian's observation of change of environment (e.g., opening of new exits), inter-individual competition, herd effect, gender differences and existence of the disabled should be taken into consideration. Concrete factors in our model are listed below. We also establish some extra rules to accomodate the crucial factors.

\begin{itemize}
   
    \item Observation of environment change: when an emergent event occurs, the evacuation personnel are likely to open extra escape doors, which aren't available in normal times. So we equip each exhibition hall with 2 doors. Suppose people are aware of this and will choose the next target which is closer to the corridor.
    \item Herd effect: individuals are influenced by the behavior of others. So we assume that people will assemble together instead of going to the door spontaneously. If there are several cells at hand for next step---say, they are of same distance to the corridor---we break the tie by allowing pedestrian to choose the step after which there are more people around him.
    \item Competition: men are usually more competitive than women, while women are more competitive than the elderly, children and the disabled. We simulate this phenomenon in the following manner. When multiple pedestrians compete for one cell at the same time, we determine who have the access based on their competitiveness values assigned randomly. People with higher competitiveness values are able to enter the next cell. 
    
\end{itemize}

\begin{figure*}
\centering
\subfigure[N=60]{
\begin{minipage}[t]{0.16\linewidth}
\centering
\includegraphics[scale=0.74]{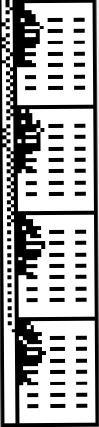}
\end{minipage}%
}%
\subfigure[N=120]{
\begin{minipage}[t]{0.16\linewidth}
\centering
\includegraphics[scale=0.74]{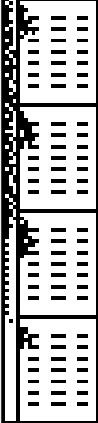}
\end{minipage}%
}%
\subfigure[N=180]{
\begin{minipage}[t]{0.16\linewidth}
\centering
\includegraphics[scale=0.74]{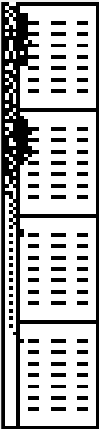}
\end{minipage}
}%
\subfigure[N=240]{
\begin{minipage}[t]{0.16\linewidth}
\centering
\includegraphics[scale=0.74]{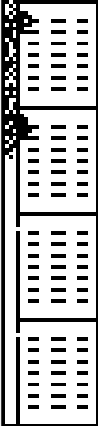}
\end{minipage}
}%
\subfigure[N=300]{
\begin{minipage}[t]{0.16\linewidth}
\centering
\includegraphics[scale=0.74]{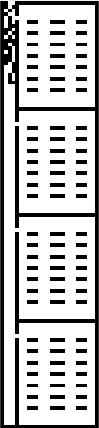}
\end{minipage}
}%
\subfigure[N=338]{
\begin{minipage}[t]{0.16\linewidth}
\centering
\includegraphics[scale=0.74]{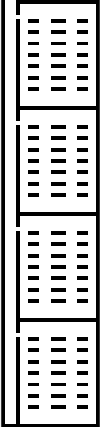}
\end{minipage}
}%
\centering
\caption{Baseline model. Each exhibition hall has one exit. No herd effect.}
\label{baseline}
\end{figure*}

\begin{figure*}
\centering
\subfigure[N=60]{
\begin{minipage}[t]{0.2\linewidth}
\centering
\includegraphics[scale=0.74]{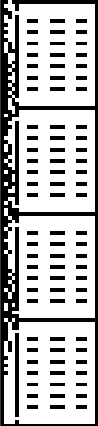}
\end{minipage}%
}%
\subfigure[N=120]{
\begin{minipage}[t]{0.2\linewidth}
\centering
\includegraphics[scale=0.74]{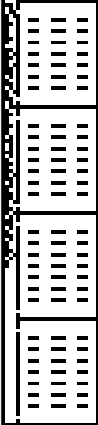}
\end{minipage}%
}%
\subfigure[N=180]{
\begin{minipage}[t]{0.2\linewidth}
\centering
\includegraphics[scale=0.74]{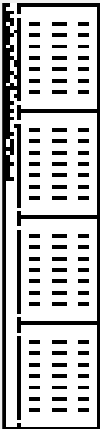}
\end{minipage}
}%
\subfigure[N=240]{
\begin{minipage}[t]{0.2\linewidth}
\centering
\includegraphics[scale=0.74]{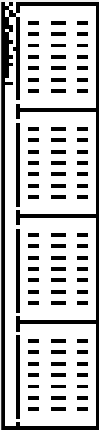}
\end{minipage}
}%
\subfigure[N=289]{
\begin{minipage}[t]{0.2\linewidth}
\centering
\includegraphics[scale=0.74]{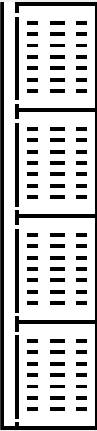}
\end{minipage}
}%
\centering
\caption{Improved model. Each exhibition hall has two exits. No herd effect.}
\label{improvewithout}
\end{figure*}

\begin{figure*}
\centering
\subfigure[N=60]{
\begin{minipage}[t]{0.16\linewidth}
\centering
\includegraphics[scale=0.74]{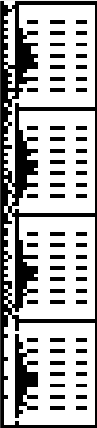}
\end{minipage}%
}%
\subfigure[N=120]{
\begin{minipage}[t]{0.16\linewidth}
\centering
\includegraphics[scale=0.74]{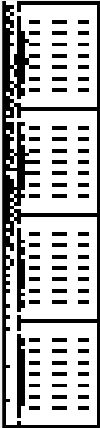}
\end{minipage}%
}%
\subfigure[N=180]{
\begin{minipage}[t]{0.16\linewidth}
\centering
\includegraphics[scale=0.74]{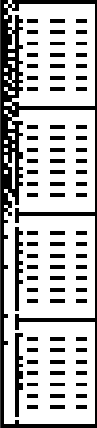}
\end{minipage}
}%
\subfigure[N=240]{
\begin{minipage}[t]{0.16\linewidth}
\centering
\includegraphics[scale=0.74]{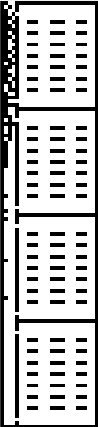}
\end{minipage}
}%
\subfigure[N=300]{
\begin{minipage}[t]{0.16\linewidth}
\centering
\includegraphics[scale=0.74]{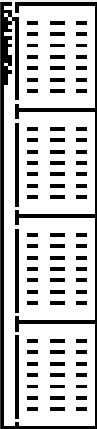}
\end{minipage}
}%
\subfigure[N=334]{
\begin{minipage}[t]{0.16\linewidth}
\centering
\includegraphics[scale=0.74]{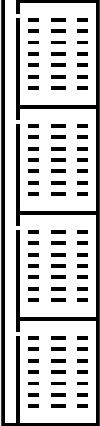}
\end{minipage}
}%
\centering
\caption{Improved model. Each exhibition hall has two exits. With herd effect.}
\label{improvewith}
\end{figure*}

\begin{figure*}
\centering
\includegraphics[scale=1]{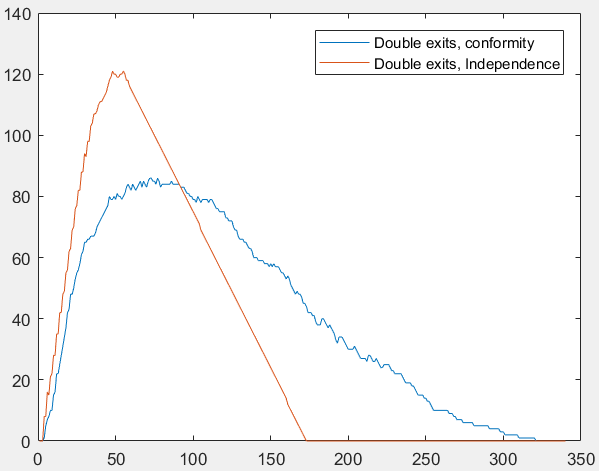}
\caption{Number of people in corridor when evacuating.}
\label{corridor}
\end{figure*}

\section{Experimental Evaluation}\label{sec:evaluation}

All experiments are conducted on an Intel machine equipped with quad-core 2.50 GHz CPU, 8GB RAM, 256GB SSD and running Windows 10.

For the baseline model, where each exhibition hall has one exit and human factors are neglected, the simulation results on MATLAB (R2017b) is shown in Figure~\ref{baseline}. It takes 338 time steps to evacuate all pedestrians.

Improved model with 2 doors for each exhibition hall finishes the evacuation process in a faster manner, which takes only 289 time steps, leading to improvement ratio of 14.50\%. The result can be seen in Figure~\ref{improvewithout}. This phenomenon highlights the importance of taking human factors into consideration. If the trapped pedestrians are unaware of the newly opened exits, the improvement in performance cannot ever be achieved. Thus taking actions to raise pedestrians' level of knowledge of the environment change can greatly contribute to effectiveness of evacuation action. Feasible measures include putting indicator lights in noticeable positions and in an outstanding manner. They can guide all pedestrian regardless of the languages they may speak. Evacuation personnel should also repeatedly inform the crowd of the newly opened exits or directly guide them toward those directions. 

However, while people's awareness of the hidden exit points can provide additional strength to an evacuation plan, their use would simultaneously cause security concerns due to the lower security postures at these exits compared with those at the four main entrances. Thus they should be carefully opened after assessing the level of threat and the requirement of evacuation efficiency of the event.  

It's observed from simulation that when there exists herd effect, the improvement in efficiency is nearly negligible, sometimes even worse (Figure~\ref{improvewith}). This is more consistent to real world scenarios. Even if more exit points are available and pedestrians are aware of this availability, the evacuation process cannot necessarily be more successful. How to reduce or even mitigate herd effect in emergencies will remain a major challenge in psychology and human factors research in the foreseeable future. One executable strategy at hand is to increase evacuation personnel to prevent herd effect and malicious competition.

Figure~\ref{corridor} shows the number of people in corridor when evacuating. Longitudinal coordinate represents the number of people in the corridor, and abscissa represents the time steps it takes to evacuate. The blue line indicates performance when there's herd effect, and the orange line when there is no herd effect. The results further indicate the harmful effect of herd effect. We can also see that the decreasing phase takes much longer time than the increasing phase. The limited number of people allowed in the corridor passage and exit becomes a bottleneck. The evacuation efficiency will be greatly affected by the width of the corridor and passing rate of its exit.

\section{Discussion}\label{sec:discussion}
\subsection{Our Recommendations}
Here we propose some feasible recommendations to improve effectiveness and efficiency of emergency evacuations.
    
We have already known that herd effect, competition and knowledge of environmental changes can affect the efficiency of crowd evacuation, so we offer some corresponding solutions.   

Firstly, before pedestrians enter crowded structures like museums and stadiums, the staff will give them electronic cards with positioning function. When an emergency occurs, the staff can track the exact location of each individual, so as to allocate more security personnel in densely populated areas, offering helpful instructions and mitigating herd effect and malicious competition. 

Secondly, the buildings should be equipped with some guiding lights. When emergency happens, the lights will guide people to less densely routes. The card can further develop the function of guiding pedestrians out of the building with LED-based route map. In this way, pedestrians will gain more knowledge of their surroundings and situations.

\subsection{Future Work}
Lack of real world data of pedestrian dynamics in emergencies has always been a challenge posed to evacuation researchers. We will tentatively apply our model to log data provided by videos in public venues, which can further showcase its advantages. 

We will also take into account more factors, not only competition but also potential cooperation. Speed difference of demo-graphical groups should also be considered.

How to make the emergency system more intelligent and automated to reduce the expense of hiring staffs is yet another question to answer.

\section{Conclusion}
In this paper, we analyze pedestrians' behavior during emergency evacuation using cellular automata simulation. Knowledge of environment change such as opening of new exit points and human factors such as competition and herd effect are considered, which is an important innovation. Experimental evaluation showcases harmful impact of herd effect and highlights the importance of informing pedestrians of newly opened exit points. Based on the results, we propose corresponding recommendations for crowded structures to improve effectiveness and efficiency of evacuation when confronting terror attacks and other emergencies. Future work is also discussed to further polish up our model.

\vspace{12pt}

\end{document}